# Vulnerability Assessment of Industrial Control System with an Improved CVSS

**He Wen**

Department of Civil and Environmental Engineering, University of Alberta, Edmonton, AB T6G 2E1, Canada

## Abstract

Cyberattacks on industrial control systems (ICS) have been drawing attention in academia. However, this has not raised adequate concerns among some industrial practitioners. Therefore, it is necessary to identify the vulnerable locations and components in the ICS and investigate the attack scenarios and techniques. This study proposes a method to assess the risk of cyberattacks on ICS with an improved Common Vulnerability Scoring System (CVSS) and applies it to a continuous stirred tank reactor (CSTR) model. The results show the physical system levels of ICS have the highest severity once cyberattacked, and controllers, workstations, and human-machine interface are the crucial components in the cyberattack and defense.

## 1. Introduction

Cyber threats on industrial control systems (ICS) have increased tremendously and received considerable attention (Kaspersky, 2022). Most of them are identified and reported as Common Vulnerabilities and Exposures (CVEs) by global stakeholders every day. Since the ICS CVEs impair the information technology (IT) and the operation technology (OT) systems, they significantly impact the facilities' process safety and functional safety (Nair et al., 2022).

Therefore, it is valuable to assess the severity of the CVEs, and currently, the Common Vulnerability Scoring System (CVSS) is a widely accepted methodology to quantify the severity, scoring 1-10 (Mell et al., 2007). The CVSS has considered the factors of attack vector (AV), attack complexity (AC), privileges required (PR), user interaction (UI), scope (S), confidentiality (C), integrity (I), availability (A), exploit code maturity (E), remediation level (RL), report confidence (RC), and environment. However, the current CVSS does not consider the safety impact of CVE. Consequently, it cannot present the severity of ICS CVE, which may trigger further accidents, not just cyber events.

For the risk assessment of the cyber system, more studies focus on the whole risk assessment (Kure et al., 2018), integrated risk analysis with multiple approaches (Leszczyna, 2021; Wen et al., 2023), system vulnerability analysis (Bolbot et al., 2019), and comprehensive defense framework (Denning, 2014). There is still a lack of vulnerability analysis of ICS and its components since ICS is an integration of IT and OT, or in other words, cyber systems and physical systems. Identifying where these vulnerabilities exist and which layers and components are more vulnerable is crucial.

Therefore, this study proposes an assessment method based on improved CVSS to locate and evaluate which parts of the ICS system are most susceptible to attack. In addition, it is applied to a continuous stirred tank reactor (CSTR) model. This study's novelty is introducing the safety factor to the CVSS system and the general results of risk assessment to provide theoretical guidance for cyber defense.

# 2. Methodology

## 2.1. Research flowchart

This study consists of two parts: methodology to identify and assess the risk due to cyberattacks on ICS and demonstration of its application to a CSTR system. The different steps involved in the study are shown in Figure 1, while a brief description is provided below.

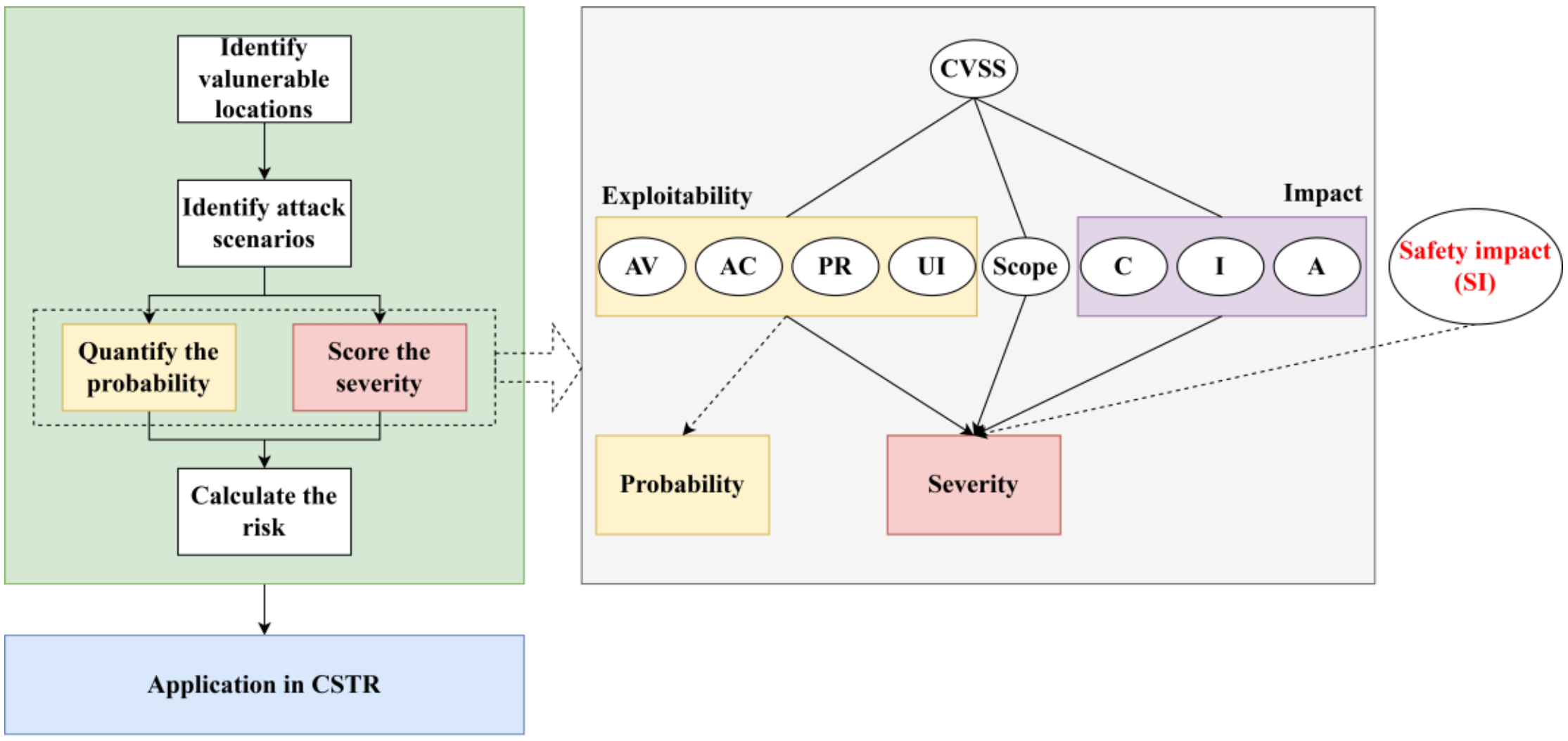

Figure 1: Details of the steps involved in the present study.

**Step 1**: Identify the vulnerable locations based on the PERA model
**Step 2**: Identify attack scenarios based on the control loop, such as components, variables, algorithms, etc.
**Step 3**: Assess the severity based on CVSS and improve it by introducing the safety factor.
**Step 4**: Quantify the probability with a proposed method based on CVSS exploitability.
**Step 5**: Calculate the risk.
**Step 6**: Apply the methodology in a CSTR model.
**Step 7**: Generate results and conclusion.

## 2.2. Identify vulnerable locations

The network is designed as a shared, interconnected platform; theoretically, all IT/OT infrastructure components could be attacked, regardless of hardware or software. According to the Purdue Enterprise Reference Architecture model (PERA) (Williams, 1994) and International Society of Automation (ISA) 95 standards (Figure 2), a non-exhaustive list of vulnerable locations is shown in Table 1.

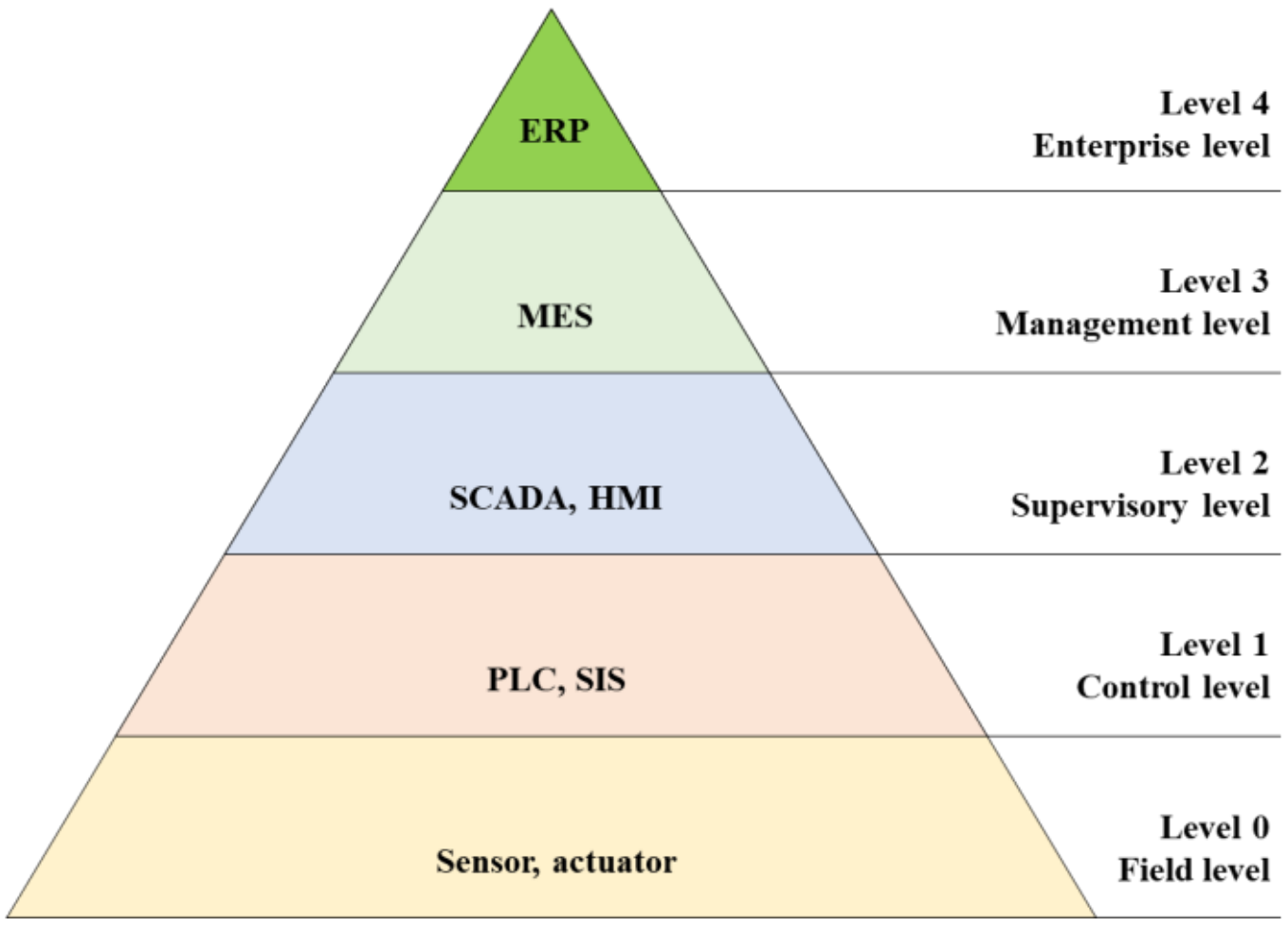

Figure 2: ICS in PERA/ISA 95 model.

Table 1: Vulnerable locations.

| Level | Vulnerable location | Similar location |
|---|---|---|
| Level 0 | Sensor | Transmitter, transducer, indicator, camera, recorder |
| | Actuator | Drive, valve, pump, fan, motor, robot arm |
| Level 1 | Programmable Logic Controller (PLC) | Intelligent Electronic Device (IED), Remote Terminal Unit (RTU) |
| | Safety instrumented system (SIS) | Protection relay |
| Level 2 | HMI | Supervisory Control and Data Acquisition (SCADA) software, Distributed Control System (DCS) software, touch screen, mobile device |
| | Control server | Supervisory controller, SCADA/DCS server |
| | Input/output server | Open Platform Communication (OPC) server |
| | Engineering workstation | Industrial computer, embedded computing device |
| | Data historian | Database |
| | Network switch | Modem, router, gateway |
| Level 3 | Management network | Manufacturing Execution System (MES), Product Lifecycle Management (PLM), Manufacturing Operations Management (MOM) |
| Level 4 | Enterprise network | Computer, server (mail server, web server), storage, E-mail, database, Enterprise Resource Planning (ERP), private cloud, network device (router, switch, modem, gateway, etc.) |

## 2.3. Identify attack scenarios

A minimal control loop model can be presented in Figure 3. A suggested frame of attack strategies or paths is shown in Table 2.

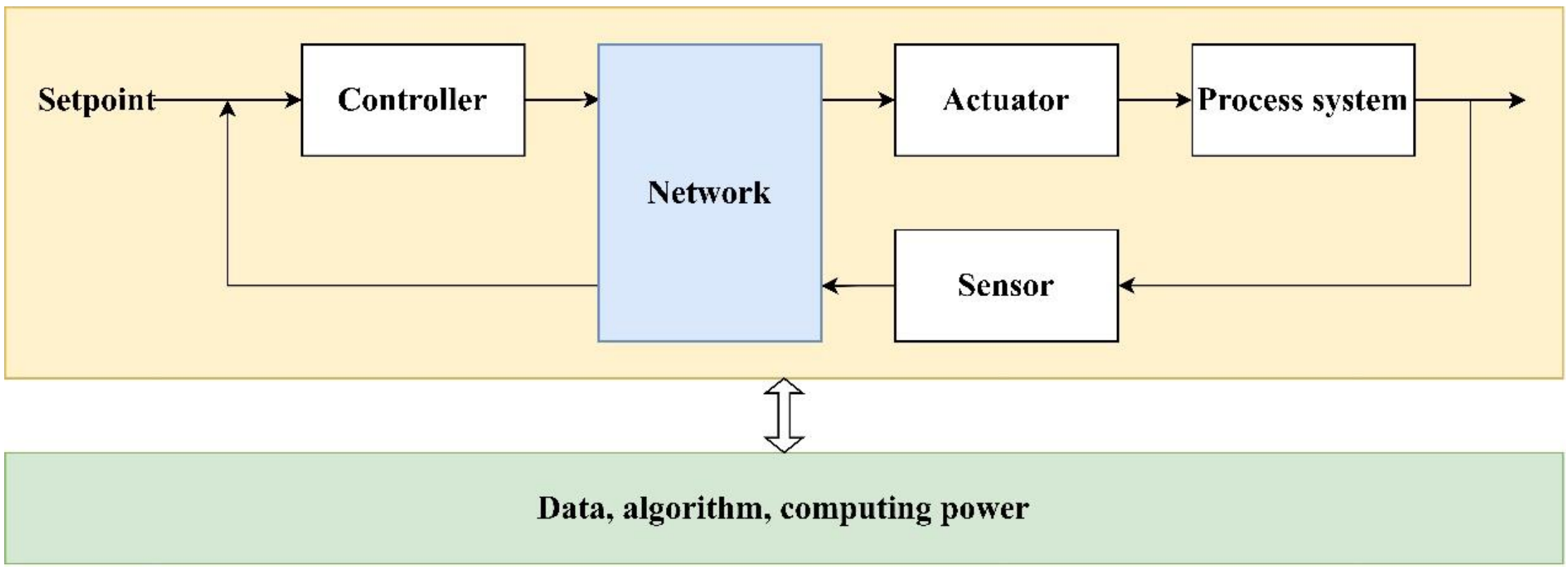

Figure 3: A minimal model of the control loop.

Table 2: Attack scenario frame in the control loop.

| Component | Parameter/vector | Attack scenarios |
|---|---|---|
| Sensor | Process variable (PV) | Manipulate the reading of sensor |
| Actuator | Manipulated variable (MV) | Manipulate MV |
| Controller | Setpoint (SP)<br>Controller parameter | Modify the setpoint<br>Modify controller algorithm, e.g., the value of Proportional, Integral, Derivative<br>Modify alarm setting |
| Network | Device | Change the configurations of the network<br>Impact or destroy the network integrity |
| | Data | Block, destroy, or modify the data/message in transmission |
| | Algorithm | Manipulate the programs |
| | Computing power | Crash the ability of servers and other computing devices |

## 2.4. Probability, severity, and risk assessment

According to CVSS V2.0, the relative likelihood or conditional probability is usually proposed (Poolsappasit et al., 2012)

$$P = 2 * AV * AC * PR \quad (1)$$

This is a widely accepted method to calculate the relative probability of vulnerability. The updated method based on CVSS V3.1 is (Zhang et al., 2017)

$$P = 2.11 * AV * AC * PR * UI \quad (2)$$

CVSS is to score the security severity; in the context of safety, safety impact on each network level should be considered (Table 3).

Table 3: Safety impact.

| Metric | Metric Value | Numerical Value |
|---|---|---|
| Safety impact (SI) | Level 0 | 1 |
| | Level 1 | 0.9 |
| | Level 2 | 0.8 |
| | Level 3 | 0.1 |
| | Level 4 | 0.05 |

An improved method to assess the severity is:

$$S = SI * f(AV, AC, PR, UI, C, I, A, Scope, SI) \quad (3)$$

And usually, the risk is

$$R = P * S \quad (4)$$

# 3. Application in CSTR

## 3.1. Identify vulnerable locations

Based on the CSTR model, the hacker may attack the following components (Figure 4).

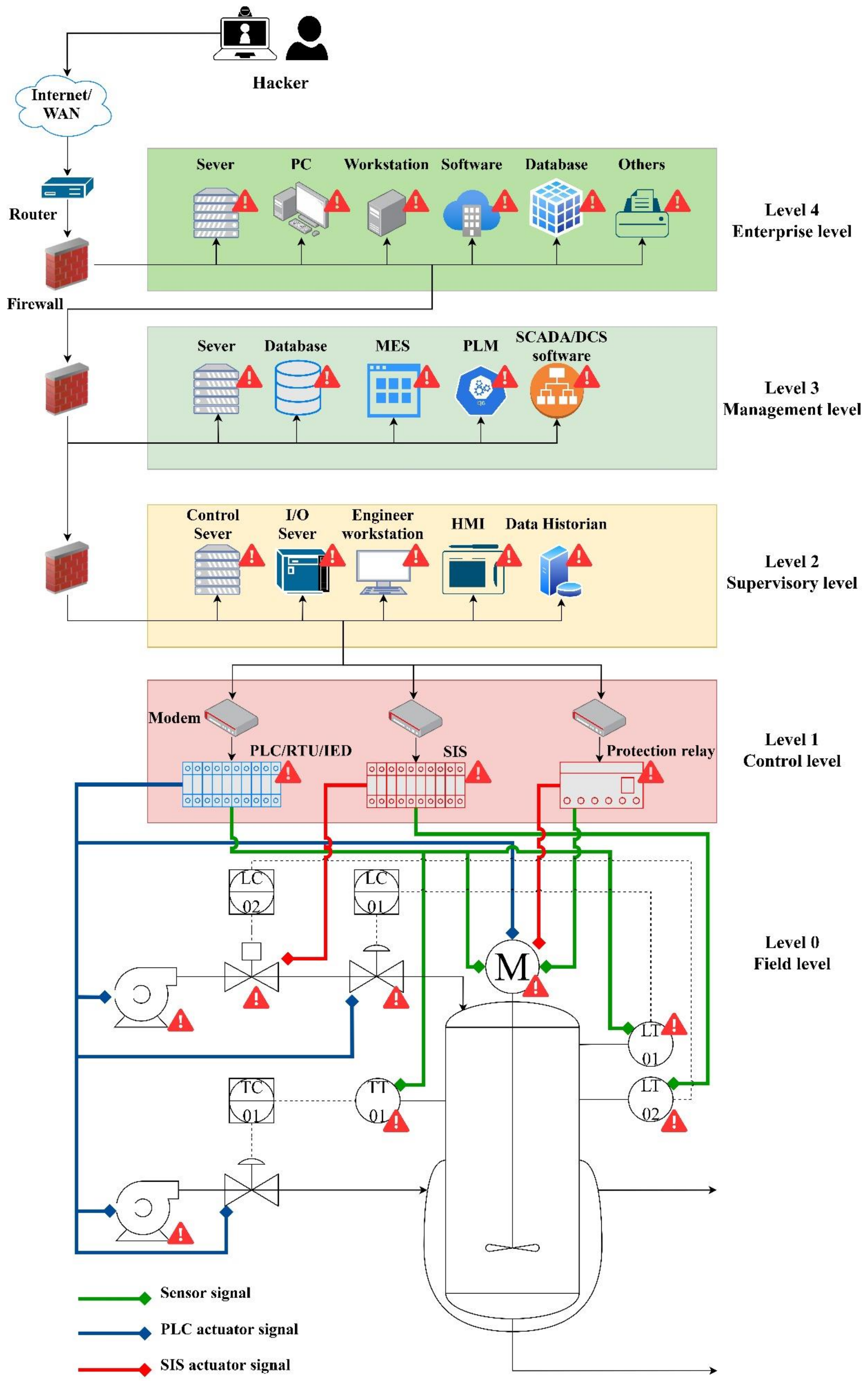

Figure 4: Vulnerable locations in CSTR system.

### 3.2. Identify attack scenarios

Thirty attack scenarios are identified, and corresponding failures and consequences are analyzed (Table 4).

Table 4: Attack scenarios in CSTR.

| No. | Level | Vulnerable location | Attack scenario in CSTR | Possible failure | Consequence |
|---|---|---|---|---|---|
| # 1 | Level 0 | Sensor | Use artificial electromagnetics in the near field to interfere with the level sensor and temperature sensor, making the readings too large or too small | Malfunction of sensors | Incorrect control of the reaction |
| # 2 | | Actuator | Use artificial electromagnetics in the near field to interfere with the speed of the pump and motor, making it too fast or too slow | Pump and motor burned or worn | Destroy the pumps, the valves, the motor, and the agitator in extreme conditions; uncontrollable reaction and secondary leakage, fire, explosion |
| # 3 | Level 1 | PLC | Brute force to change I/O point values of PLC to change the speed of the pump or the motor. | Pump and motor worn | Destroy the pumps, the valves, the motor, and the agitator in extreme conditions; uncontrollable reaction and secondary leakage, fire, explosion |
| # 4 | | PLC | Use worms to infect one PLC, then the worm locates and infects other PLCs; all infected PLCs implement an endless access request to the control sever or the engineering workstation | The control server or engineering workstation stops the service | No control action and no state change |
| # 5 | | PLC | Use malware to infect PLC and block command messages, block reporting messages, block serial COM, or destroy data directly | No command from PLC | No control action and no state change |
| # 6 | | PLC | Manipulate the level/temperature sensor data or data to the valve/pump/motor to cause the PLC to control the reaction incorrectly | Undesired command from PLC | Destroy the pumps, the valves, the motor, and the agitator in extreme conditions; uncontrollable reaction and secondary leakage, fire, explosion |
| # 7 | | PLC | Manipulate the program/task of the PLC | Undesired command from PLC | Destroy the pumps, the valves, the motor, and the agitator in extreme conditions; uncontrollable reaction |
| | | | | | and secondary leakage, fire, explosion |
| # 8 | | PLC | Manipulate the status of PLC by changing operating mode, such as device restart/shutdown | Pump and motor burned or worn | Destroy the pumps, the valves, the motor, and the agitator in extreme conditions; uncontrollable reaction and secondary leakage, fire, explosion |
| # 9 | | PLC | Manipulate the view or I/O image of the PLC | The operator may perform wrong action | Incorrect control of the reaction |
| # 10 | | PLC | Modify the setpoint of level/temperature | Wrong command from PLC | Destroy the pumps, the valves, the motor, and the agitator in extreme conditions; uncontrollable reaction and secondary leakage, fire, explosion |
| # 11 | | PLC | Modify the alarm setting or disables the alarm program | Alarm failure | Lose alarm and protection |
| # 12 | | SIS | Intentionally activate SIS to shut down the process system | SIS malfunction | Uncontrollable shutdown |
| # 13 | | SIS | Reprogram SIS or block the activation to lose the protection function | SIS failure | Lose function and protection |
| # 14 | | SIS | Brute force to activate and deactivate the SIS frequently | SIS failure | Destroy the valve, motor, or agitator |
| # 15 | Level 2 | HMI | Control the Command-Line Interface or Graphical User Interface to execute a command | Undesired command from PLC | Uncontrollable reaction and secondary leakage, fire, explosion |
| # 16 | | HMI | Manipulate the data or view of HMI | The operator may perform wrong action | Incorrect control of the reaction |
| # 17 | | HMI | Modify the setpoint of level/temperature on HMI | Wrong command to PLC | Uncontrollable reaction and secondary leakage, fire, explosion |
| # 18 | | HMI | Wipe the data or image to make a loss of view or service stop | No command to PLC | No control action and no state change |
| # 19 | | HMI | Modify the alarm setting or remove the indicator on the host | Alarm failure | Lose alarm and protection |
| # 20 | | Control server | Set up a rogue master to leverage control server functions to communicate with outstations | Wrong command to PLC | Uncontrollable reaction and secondary leakage, fire, explosion |
| # 21 | | Control server | Conduct a Distributed Denial-of-Service (DDoS) attack on the server | No command to PLC | No control action and no state change |
| # 22 | | Input/output server | Use malware to block data transfer to PLC | No command to PLC | No control action and no state change |
| # 23 | | Input/output server | Access into industrial environments through systems exposed directly to the internet for remote access | Undesired command to PLC | Uncontrollable reaction and secondary leakage, fire, explosion |
| # 24 | | Engineering workstation | Use malware or malicious code to infect project files | Data loss affecting the performance of PLC | Incorrect control of the reaction |
| # 25 | | Engineering workstation | Use a spear phishing attachment as a form of a social engineering attack against specific targets | The operator may perform wrong action | Incorrect control of the reaction |
| # 26 | | Data historian | Collect data | Data breach | Undesired consequence |
| # 27 | | Data historian | Destruct data | Data loss affecting the performance of PLC | Incorrect control of the reaction |
| # 28 | | Network switch | ARP spoofing on Internet facilities | Internet outage or failure to transfer data | No control action and no state change |
| # 29 | Level 3 | Management network | Attack the management systems with malware, ransomware, phishing, DDoS | Data breach, Internet outage | Business interruption or direct economic loss |
| # 30 | Level 4 | Enterprise network | Attack the enterprise systems/software with malware, ransomware, phishing, DDoS | Data breach, Internet outage | Business interruption or direct economic loss |

### 3.3. Probability, severity, and risk assessment

CVSS is to score the security severity; in the context of safety, safety impact on each network level should be considered.
After analyzing the metrics of each scenario, the severity, probability, and risk could be calculated with Eq. 2, Eq, 3, and Eq. 4 (Table 5).
Table 5: Severity, probability, and risk

| No. | AV | AC | PR | UI | Confidentiality | Integrity | Availability | Scope | Probability | Severity | Risk |
|---|---|---|---|---|---|---|---|---|---|---|---|
| # 1 | Physical | HIGH | High | None | None | Low | Low | Unchanged | 0.13 | 2.70 | 0.36 |
| # 2 | Physical | HIGH | High | None | None | None | High | Unchanged | 0.13 | 3.80 | 0.51 |
| # 3 | Adjacent | HIGH | High | None | None | Low | Low | Unchanged | 0.13 | 2.79 | 0.37 |
| # 4 | Adjacent | HIGH | High | None | None | Low | High | Unchanged | 0.13 | 4.32 | 0.57 |
| # 5 | Adjacent | HIGH | High | None | High | High | High | Unchanged | 0.13 | 5.76 | 0.76 |
| # 6 | Adjacent | HIGH | High | None | High | High | Low | Unchanged | 0.13 | 5.40 | 0.71 |
| # 7 | Adjacent | HIGH | High | None | High | High | High | Unchanged | 0.13 | 5.76 | 0.76 |
| # 8 | Adjacent | HIGH | High | None | Low | Low | High | Unchanged | 0.13 | 4.77 | 0.63 |
| # 9 | Adjacent | HIGH | High | Required | Low | High | Low | Unchanged | 0.10 | 4.59 | 0.44 |
| # 10 | Adjacent | HIGH | High | None | Low | High | Low | Unchanged | 0.13 | 4.77 | 0.63 |
| # 11 | Adjacent | HIGH | High | None | Low | Low | High | Unchanged | 0.13 | 4.77 | 0.63 |
| # 12 | Adjacent | HIGH | High | None | None | None | High | Unchanged | 0.13 | 3.78 | 0.50 |
| # 13 | Adjacent | HIGH | High | None | None | None | High | Unchanged | 0.13 | 3.78 | 0.50 |
| # 14 | Adjacent | HIGH | High | None | None | None | High | Unchanged | 0.13 | 3.78 | 0.50 |
| # 15 | Network | HIGH | High | None | High | High | High | Unchanged | 0.18 | 5.28 | 0.96 |
| # 16 | Network | HIGH | High | Required | Low | High | Low | Unchanged | 0.13 | 4.24 | 0.56 |
| # 17 | Network | HIGH | High | None | Low | High | High | Unchanged | 0.18 | 4.96 | 0.90 |
| # 18 | Network | HIGH | High | None | Low | High | Low | Unchanged | 0.18 | 4.40 | 0.80 |
| # 19 | Network | HIGH | High | None | Low | High | High | Unchanged | 0.18 | 4.96 | 0.90 |
| # 20 | Network | HIGH | High | None | Low | Low | Low | Unchanged | 0.18 | 3.28 | 0.59 |
| # 21 | Network | HIGH | High | None | None | None | High | Unchanged | 0.18 | 3.52 | 0.64 |
| # 22 | Adjacent | HIGH | High | Required | None | High | Low | Unchanged | 0.10 | 3.68 | 0.35 |
| # 23 | Network | LOW | High | None | High | None | High | Unchanged | 0.32 | 5.20 | 1.65 |
| # 24 | Network | LOW | Low | Required | Low | High | Low | Unchanged | 0.53 | 5.44 | 2.89 |
| # 25 | Network | HIGH | None | Required | None | Low | Low | Unchanged | 0.73 | 4.32 | 3.14 |
| # 26 | Network | HIGH | High | None | High | High | None | Unchanged | 0.18 | 4.72 | 0.85 |
| # 27 | Network | HIGH | High | None | High | High | None | Unchanged | 0.18 | 4.72 | 0.85 |
| # 28 | Network | HIGH | High | None | None | None | High | Unchanged | 0.18 | 3.52 | 0.64 |
| # 29 | Network | LOW | Low | None | High | High | High | Unchanged | 0.73 | 0.88 | 0.64 |
| # 30 | Network | LOW | Low | None | High | High | High | Unchanged | 0.73 | 0.44 | 0.32 |

For different system layers and major components, the summary is presented in Table 6.

Table 6: Averaged severity and risk for each level and each vulnerable location.

| Level | Severity | Risk | Vulnerable location | Severity | Risk |
|---|---|---|---|---|---|
| Level 0 | 3.25 | 0.44 | Sensor | 2.70 | 0.36 |
| | | | Actuator | 3.80 | 0.51 |
| Level 1 | 4.65 | 0.60 | PLC | 4.94 | 0.63 |
| | | | SIS | 3.78 | 0.50 |
| Level 2 | 4.36 | 1.1 | HMI | 4.77 | 0.82 |
| | | | Control server | 3.40 | 0.62 |
| | | | Input/output server | 4.44 | 1.00 |
| | | | Engineering workstation | 4.88 | 3.02 |
| | | | Data historian | 4.12 | 0.75 |
| | | | Network switch | 3.52 | 0.64 |
| Level 3 | 0.88 | 0.64 | Management network | 0.88 | 0.64 |
| Level | 0.44 | 0.32 | Enterprise network | 0.44 | 0.32 |

According to the statistics, it can be concluded that Levels 1 and 2 rank the highest severity of cyberattacks on safety impact, and Level 4 has no significant safety impact on physical systems (severity ranking: Level 1>Level 2>Level 0>Level 3>Level 4). On the other hand, Level 2 shows the highest risk since Level 1 is much more difficult to attack successfully than Level 2 (risk ranking: Level 2>Level 3>Level 1>Level 0>Level 4). Specifically, PLC, engineering workstation, and HMI rank the highest severity; engineering workstation ranks the highest risk, followed by I/O sever and HMI.

## 4. Conclusions

This study develops a generic approach to assess the cyber risk of ICS systems and components based on an improved CVSS. A case study on CSTR has also been conducted. The results show ICS's vulnerable locations and parts and present the common attack scenarios and techniques.

This study is a reminder to industrial practitioners who used to consider that cyberattacks only touch network systems and do not impair physical systems. On the contrary, the physical system is the ultimate attack target against ICS, which will cause the most severe damage. In terms of risk, PLC, engineering workstation, and HMI are the critical objects that need to be protected since they control the stable operation of ICS.

## References

Bolbot, V., Theotokatos, G., Bujorianu, L. M., Boulougouris, E., & Vassalos, D. (2019). Vulnerabilities and safety assurance methods in Cyber-Physical Systems: A comprehensive review. In *Reliability Engineering and System Safety* (Vol. 182, pp. 179–193). https://doi.org/10.1016/j.ress.2018.09.004

Denning, D. E. (2014). Framework and principles for active cyber defense. *Computers*

*and Security*, *40*, 108–113. https://doi.org/10.1016/j.cose.2013.11.004

Kaspersky. (2022). Threat landscape for industrial automation systems: Statistics for H1 2022. In *AO Kaspersky Lab*.

Kure, H. I., Islam, S., & Razzaque, M. A. (2018). An integrated cyber security risk management approach for a cyber-physical system. *Applied Sciences (Switzerland)*, *8*(6). https://doi.org/10.3390/app8060898

Leszczyna, R. (2021). Review of cybersecurity assessment methods: Applicability perspective. *Computers and Security*, *108*. https://doi.org/10.1016/j.cose.2021.102376

Mell, P., Scarfone, K., & Romanosky, S. (2007). A complete guide to the common vulnerability scoring system version 2.0. *Published by FIRST-Forum of Incident Response and Security Teams*, *1*, 23.

Nair, A., Ray, A., Reddy, L., & Marali, M. (2022). Mapping of CVE-ID to Tactic for Comprehensive Vulnerability Management of ICS. In *Inventive Communication and Computational Technologies: Proceedings of ICICCT 2022* (pp. 559–571). Springer.

Poolsappasit, N., Dewri, R., & Ray, I. (2012). Dynamic security risk management using Bayesian attack graphs. *IEEE Transactions on Dependable and Secure Computing*, *9*(1), 61–74. https://doi.org/10.1109/TDSC.2011.34

Wen, H., Khan, F., Ahmed, S., Imtiaz, S., & Pistikopoulos, S. (2023). Risk assessment of human-automation conflict under cyberattacks in process systems. *Computers & Chemical Engineering*, *172*, 108175. https://doi.org/https://doi.org/10.1016/j.compchemeng.2023.108175

Williams, T. J. (1994). The Purdue Enterprise Reference Architecture and Methodology (PERA). *Computers in Industry*, *24*(2–3), 141–158.

Zhang, H., Lou, F., Fu, Y., & Tian, Z. (2017). A conditional probability computation method for vulnerability exploitation based on CVSS. *Proceedings - 2017 IEEE 2nd International Conference on Data Science in Cyberspace, DSC 2017*, 238–241. https://doi.org/10.1109/DSC.2017.33